\documentclass[aps,prb,preprint,showpacs,amsmath,amssymb,amsfonts]{revtex4-1}

\usepackage{graphicx}
\usepackage{color}

\begin{document}

\title{Quasinormal modes expansions for nanoresonators made of absorbing dielectric materials: study of the role of static modes}

\author{Christophe Sauvan}\email[Corresponding author: ]{christophe.sauvan@institutoptique.fr}
\affiliation{Universit\'e Paris-Saclay, Institut d'Optique Graduate School, CNRS, Laboratoire Charles Fabry, 91127, Palaiseau, France}

\date{\today}

\begin{abstract}
The interaction of light with photonic resonators is determined by the eigenmodes of the system. Modal theories based on quasinormal modes provide a natural tool to calculate and understand light scattering by nanoresonators. We show that, in the case of resonators made of absorbing dielectric materials, eigenmodes with zero eigenfrequency (static modes) play a key role in the modal formalism. The excitation of static modes builds a non-resonant contribution to the modal expansion of the scattered field. This non-resonant term plays a crucial physical role since it largely contributes to the off-resonance signal to which resonances are added in amplitude, possibly leading to interference phenomena and Fano resonances. By considering light scattering by a silicon nanosphere, we quantify the impact of static modes. This study shows that the importance of static modes is not just formal. Modal expansions without static modes reconstruct an incorrect internal field and incorrect extinction and absorption cross-sections. Static modes are of prime importance in an expansion truncated to only a few modes.
\end{abstract}

\maketitle

%%%%%%%%%%%%%%%%%%%%%%%%%%  body  %%%%%%%%%%%%%%%%%%%%%%%%%%
\section{Introduction}

Nanoresonators play an important and growing role in numerous applications of optical science, from lasers~\cite{Fainman12,Hill14} to high-performance sensors~\cite{Giessen11,SensorDielCR16,GiessenCR17}. Like other resonance phenomena in wave physics, light scattering by a resonant photonic system is largely determined by the eigenmodes of the structure. When an optical resonator is driven by an incident beam with a frequency close to an eigenfrequency of the system, the corresponding mode is excited and it results in the appearance of a resonance in the spectrum of the resonator response.

Nanoresonators are open systems often made of absorbing and dispersive materials and, as such, are described by non-Hermitian operators. Their eigenmodes are usually called quasinormal modes, leaky modes, or resonant states~\cite{QNMReview}. We refer to them in the following as quasinormal modes, abbreviated in QNMs, or simply modes. Owing to the non-Hermitian character of the system and the spectral dispersion of the constituent materials, the development of QNM theories for optical resonators is a non-trivial task, which, after initial works in the 90's~\cite{YoungPRA94,YoungJOSAB96}, has recently received much attention~\cite{QNMReview,HughesOL12,SauvanPRL13,BaiOE13,MuljarovPRA14,MuljarovPRB16,MansuPRA17,WeiPRB2018,FresnelOL18,GralakOSA18,MuljarovPRA18,BurgerPRA18,WeissPRB18,BonodPRB18,DelftOE20,FresnelOE20,BxJOSAA20}. In simple words, QNM theories translate a physical observation into mathematical terms by describing the system response with a modal expansion, or decomposition, which is nothing more than a sum of resonant terms. Each term corresponds to the excitation of a mode and resonates in a frequency range close to the mode eigenfrequency~\cite{QNMReview}.

Resonant terms are not always sufficient to fully determine the optical properties of a resonator driven by an external excitation. On the one hand, off-resonance signals cannot always be predicted by simply adding the resonant contributions of a few modes whose eigenfrequencies are close to the spectral range of interest. On the other hand, the peculiar spectral shape of a Fano resonance results from the superposition of a resonant channel (the excitation of a mode) and a non-resonant channel (often referred to as a continuum)~\cite{Fano61,Kivshar17}. In both cases, a significant role is played by contributions to the electromagnetic field \emph{that do not resonate in the spectral range of interest}. However, the origin of these contributions is not always clear. Do they come from the excitation of one (or a few) mode with a very low quality factor whose response seems flat in a given spectral window? From the excitation of a large number of modes whose eigenfrequencies are located far from the spectral range of interest? From a purely non-resonant term (i.e., that does not resonate whatever the spectral range) that adds to the resonant contributions of the modes? From a mix of these effects? Let us emphasize that, in the following, we use the adjective "non-resonant" to characterize something that is spectrally smooth whatever the spectral range, i.e., that does not resonate, even outside the spectral range of interest.

These questions have been explicitly or implicitly addressed by recent works on QNM theory, but no clear answers emerge. On the one hand, most approaches leading to a modal theory do not introduce an explicit non-resonant term and express the scattered field or the Green tensor as a sum of modal contributions~\cite{SauvanPRL13,BaiOE13,MuljarovPRA14,MuljarovPRB16,MansuPRA17,WeiPRB2018,FresnelOL18,GralakOSA18,MuljarovPRA18,FresnelOE20,BxJOSAA20,YoungPRA94,YoungJOSAB96}. Modes are the only ingredients of these approaches, which are based on the assumption that the QNMs supported by the resonator forms a complete set. Note that the completeness has been demonstrated for simple geometries such as slabs and spheres~\cite{MansuPRA17,GralakOSA18,YoungPRA94,YoungJOSAB96} but not for more complex systems. Within these formalisms, if a non-resonant contribution to the scattered field shall exist, it should somehow be included in the modal expansion.

On the other hand, a few works introduce an explicit non-resonant contribution in addition to the sum of resonant terms, each corresponding to the excitation of a mode~\cite{BurgerPRA18,WeissPRB18,BonodPRB18,BonodJOSAB19}. In~\cite{BurgerPRA18}, the total field is decomposed into resonant and non-resonant components whose calculation is based on Riesz projections. The non-resonant components, mixed with the contributions from modes outside the spectral range of interest, are calculated by contour integration in the complex frequency plane, see Eqs. (3)-(5) in~\cite{BurgerPRA18}. At each integration point, the total electric field is rigorously calculated with a finite-element method. In~\cite{WeissPRB18}, T. Weiss \emph{et al.} derive a modal expansion for the scattering matrix of periodic resonant systems and introduce an additional non-resonant term called background term, see Eq. (23) in~\cite{WeissPRB18}. Like in~\cite{BurgerPRA18}, the non-resonant background is mixed with the influence of modes outside the spectral range of interest. In practice, it is taken into account by a cubic fit to the exact spectrum calculated at a few equidistant frequencies~\cite{WeissPRB18}. In~\cite{BonodPRB18,BonodJOSAB19}, R. Colom \emph{et al.} study light scattering by dielectric (non-absorbing) spheres with a modal analysis. They show that a modal expansion of the scattering matrix contains a non-resonant term while a modal expansion of the internal field does not. The derivation is mainly based on causality and energy conservation principles~\cite{BonodPRB18}. In this case of a spherical lossless scatterer, the non-resonant term takes a simple closed-form expression~\cite{BonodPRB18,BonodJOSAB19}.

This brief review of recent works on QNM theory in nanophotonics shows that two main approaches providing accurate results coexist: the scattered field is either decomposed into a sum of modal contributions or into a sum of modal contributions plus a non-resonant term. In the present article, we bridge the gap between these seemingly contradictory approaches. Starting from a QNM formalism with only modal contributions, we show that, in the case of resonators made of absorbing dielectric materials, eigenmodes with zero eigenfrequency (static modes) play a key role in the modal theory. Indeed, unlike other QNMs with a non-zero eigenfrequency, static modes provide a non-resonant contribution to the modal expansion of the scattered field. In other words, they are not poles of the system. This result demonstrates that, provided that static modes are properly taken into account, a modal expansion contains a non-resonant term. The existence of static modes and their role in QNM perturbation theories (i.e., the description of a perturbed system with the QNMs of an unperturbed resonator) has been discussed in previous works, see~\cite{MuljarovPRA14,MuljarovPRA19,WeiPRL20}. However, to the best of our knowledge, the impact of static modes in QNM theory of light scattering has not been quantified and their role in the construction of a non-resonant contribution to the scattered field has not been discussed.

We start Section 2 by a brief reminder of the QNM theory used in the rest of the article. Then we show that static modes (modes with zero eigenfrequency) contribute non-resonantly to the scattered field. In Sections 3 and 4, we illustrate with concrete examples that the importance of static modes is not only formal. We show that, in the case of resonators made of dielectric (non-metallic) materials, QNM expansions without static modes lead to significant errors and that static modes are of prime importance in an expansion truncated to only a few modes. Static modes have also a crucial physical role since they largely contribute to the background (off-resonance) signal to which resonances are added in amplitude, possibly leading to interference phenomena. We first consider in Section 3 the simple case of a lossless sphere with a constant and real refractive index. Section 4 is devoted to the case of sphere made of an absorbing and dispersive material whose relative permittivity is given by a Lorentz model. We have limited our study to spheres because they are the only three-dimensional system where the convergence of the modal expansion can be tested down to a very low error level. Indeed, Mie theory provides an analytical calculation of the modes (static and non-static) and an exact reference calculation of the field.

\section{Static modes and their non-resonant contribution to the modal expansion}

In this Section, we first recall important results of quasinormal modes theory. More details can be found in~\cite{QNMReview,WeiPRB2018,FresnelOE20} and references therein. Then, we focus on the role of static modes (modes with zero eigenfrequency) in the modal expansion and we show that, in contrast to other QNMs, these modes contribute non-resonantly to the scattered field. As a consequence, the scattered field calculated from a QNM expansion is not only composed of resonant terms but also of non-resonant terms. Their role is extremely important since a modal expansion without a correct non-resonant contribution is not able to accurately predict all the important features of the scattering problem.

\subsection{Quasinormal modes theory: a brief reminder}

We consider a nanoresonator characterized by a permittivity distribution $\varepsilon({\bf r},\omega)$ and illuminated by an incident beam with an electric field ${\bf E}_i({\bf r})$. For the sake of simplicity, we recall below the main elements of QNM theory for non-magnetic materials with $\mu({\bf r},\omega) = \mu_0$, but the theory equally applies to magnetic materials with $\mu \neq \mu_0$. In the scattered-field formulation, the permittivity of the complete system is decomposed as $\varepsilon({\bf r},\omega) = \varepsilon_b({\bf r},\omega) + \Delta\varepsilon({\bf r},\omega)$, where $\varepsilon_b({\bf r},\omega)$ is the permittivity of the background that surrounds the resonator and $\Delta\varepsilon({\bf r},\omega)$ is null everywhere except inside the resonator. Note that the background is not necessarily a homogeneous medium; the simplest non-homogeneous background being the interface between an incident medium and a substrate.

Without any loss of generality, the total electric field can be decomposed as

\begin{equation}\label{eq:Et}
    {\bf E}({\bf r},\omega) = {\bf E}_b({\bf r},\omega) + {\bf E}_s({\bf r},\omega) ,
\end{equation}

\noindent where the background field ${\bf E}_b$ is the field in the absence of the resonator (i.e., for $\Delta\varepsilon = 0$) and ${\bf E}_s$ is the scattered field, which results from the presence of $\Delta\varepsilon({\bf r},\omega)$. Note that ${\bf E}_b = {\bf E}_i$ if the background is a homogeneous medium and ${\bf E}_b \neq {\bf E}_i$ otherwise.

The QNM formalism consists in decomposing the scattered field into a sum of modes~\cite{QNMReview,WeiPRB2018,FresnelOE20},

\begin{equation}\label{eq:Es}
    {\bf E}_s({\bf r},\omega) = \sum_m \alpha_m(\omega) \tilde{{\bf E}}_m({\bf r}) ,
\end{equation}

\noindent where the contribution of each mode is weighted by an excitation coefficient $\alpha_m(\omega)$, which depends solely on the frequency. Each QNM is characterized by a complex eigenfrequency $\tilde{\omega}_m$ and a field distribution $[\tilde{{\bf E}}_m({\bf r}),\tilde{{\bf H}}_m({\bf r})]$, which is solution of source-free Maxwell's equations,

\begin{subequations}\label{eq:Maxwell}
\begin{align}
    \nabla \times \tilde{{\bf E}}_m & = i \tilde{\omega}_m \mu_0 \tilde{{\bf H}}_m \,, \\
    \nabla \times \tilde{{\bf H}}_m & = -i \tilde{\omega}_m \varepsilon(\tilde{\omega}_m) \tilde{{\bf E}}_m \,,
\end{align}
\end{subequations}

\noindent with outgoing-waves boundary conditions~\cite{QNMReview,QNMSolver}.

For a system made of dispersive materials whose permittivity $\varepsilon(\omega)$ is described by a rational function of the frequency, it has recently been shown that the closed-form expression of the excitation coefficients is not unique~\cite{FresnelOE20,BxJOSAA20}. The non-uniqueness of the excitation coefficients is related to the over-completeness of the QNMs set and to the non-uniqueness of the linearization of Maxwell's equations in dispersive media~\cite{FresnelOE20,BxJOSAA20}. In this work, we use the following expression for $\alpha_m(\omega)$, which is obtained by linearizing the equations satisfied by the scattered field with a polarization and a current density as auxiliary fields\cite{BxJOSAA20},

\begin{equation}\label{eq:alpha}
    \alpha_m(\omega) = -\frac{\omega}{\omega - \tilde{\omega}_m} \iiint_V \Delta\varepsilon({\bf r},\omega) {\bf E}_b({\bf r},\omega) \cdot \tilde{{\bf E}}_m({\bf r}) d^3{\bf r} = \frac{\omega}{\omega - \tilde{\omega}_m} a_m(\omega) \,,
\end{equation}

\noindent with $V$ the volume inside which $\Delta\varepsilon \neq 0$ and $a_m(\omega) = -\iiint_V \Delta\varepsilon({\bf r},\omega) {\bf E}_b({\bf r},\omega) \cdot \tilde{{\bf E}}_m({\bf r}) d^3{\bf r}$. Equation~\eqref{eq:alpha} holds provided that the QNMs are normalized as follows~\cite{QNMReview,SauvanPRL13,BaiOE13,QNMSolver}:

\begin{equation}\label{eq:Norm}
    \iiint_\Omega \left [ \tilde{{\bf E}}_m \cdot \frac{\partial\omega\varepsilon(\omega)}{\partial\omega} \tilde{{\bf E}}_m - \tilde{{\bf H}}_m \cdot \frac{\partial\omega\mu(\omega)}{\partial\omega} \tilde{{\bf H}}_m  \right ] d^3{\bf r} = 1 \,,
\end{equation}

\noindent where the derivatives are taken at the QNM frequency $\omega = \tilde{\omega}_m$. Because of the divergence of the QNM field as ${\bf r} \rightarrow 0$, the integration domain $\Omega$ is defined in a complex coordinate space in order to damp the exponential divergence~\cite{QNMReview,SauvanPRL13}. In practice, complex coordinates can be numerically implemented with perfectly-matched layers~\cite{Chew94,Nicolet07}.

According to Eq.~\eqref{eq:alpha}, QNMs are poles of the scattered field since the excitation coefficients vary as $1/(\omega - \tilde{\omega}_m)$. As a consequence, the scattered field seems to be a sum of resonant contributions without non-resonant term, see Eq.~\eqref{eq:Es}. On the other hand, the total field is the sum of a single non-resonant term (the background field) and resonant terms (the QNMs), see Eq.~\eqref{eq:Et}. We will see in the following that this conclusion is erroneous if the system supports static modes with $\tilde{\omega}_m=0$. The latter provide in fact a non-resonant contribution to the scattered field.

\subsection{Static modes and their non-resonant contribution}
\label{sec:static_modes}

Static modes are solutions of source-free Maxwell's equations, see Eq.~\eqref{eq:Maxwell}, for $\tilde{\omega}_m=0$. They are also referred to as longitudinal modes since the curl of their electric (or magnetic) field is equal to zero~\cite{MuljarovPRA14}. Two distinct groups of static modes exist. Electric static modes have zero magnetic field, $\tilde{{\bf H}}_m^{st} = 0$, and a curl-free electric field, $\nabla \times \tilde{{\bf E}}_m^{st} = 0$, whereas magnetic static modes have zero electric field, $\tilde{{\bf E}}_m^{st} = 0$, and a curl-free magnetic field, $\nabla \times \tilde{{\bf H}}_m^{st} = 0$. Since we consider in this work non-magnetic resonators ($\mu = \mu_0$ and $\Delta\mu = 0$), magnetic static modes cannot be excited by an incident beam. Therefore, we discuss hereafter only electric static modes.

It is noteworthy that the existence of two separate families of curl-free static modes is true if, and only if, the source-free Maxwell's equations reduce to

\begin{subequations}\label{eq:Maxwell_sta}
\begin{align}
    \nabla \times \tilde{{\bf E}}_m^{st} & = 0 \,, \\
    \nabla \times \tilde{{\bf H}}_m^{st} & = 0 \,,
\end{align}
\end{subequations}

\noindent as $\tilde{\omega}_m \rightarrow 0$. In other words, the behavior of the permittivity in the static regime has to be such that

\begin{equation}\label{eq:staticlimit}
    \lim_{\omega \rightarrow 0} \omega \varepsilon(\omega) = 0 .
\end{equation}

\noindent This condition is fulfilled by materials whose permittivity is described by a Lorentz model, but not by metals described by a Drude model. Indeed, $\omega = 0$ is a pole of the Drude permittivity and $\omega \varepsilon(\omega)$ tends towards a finite non-zero value as $\omega \rightarrow 0$. In the latter case, the existence and the nature of static modes is a more complex issue since electric and magnetic fields are not decoupled as in usual electrostatics and magnetostatics.

\medskip

We consider in the following resonators made of materials whose permittivity fulfills Eq.~\eqref{eq:staticlimit}. Since the eigenfrequency of an electric static mode is null, $\tilde{\omega}_m^{st} = 0$, its excitation coefficient becomes

\begin{equation}\label{eq:alpha_sta}
    \alpha_m^{st}(\omega) = \frac{\omega}{\omega} a_m^{st}(\omega) = a_m^{st}(\omega) \,,
\end{equation}

\noindent with $a_m^{st}(\omega) = -\iiint_V \Delta\varepsilon({\bf r},\omega) {\bf E}_b({\bf r},\omega) \cdot \tilde{{\bf E}}_m^{st}({\bf r}) d^3{\bf r}$. Since $\tilde{{\bf H}}_m^{st} = 0$, the expression of $\alpha_m^{st}(\omega)$ holds provided that the static mode is normalized as follows,

\begin{equation}\label{eq:Norm_sta}
    \iiint_\Omega \tilde{{\bf E}}_m^{st} \cdot \frac{\partial\omega\varepsilon(\omega)}{\partial\omega} \tilde{{\bf E}}_m^{st}  d^3{\bf r} = 1 \,.
\end{equation}

\noindent The derivative is taken at the eigenfrequency $\omega = \tilde{\omega}_m^{st} = 0$.

According to Eq.~\eqref{eq:alpha_sta}, the eigenfrequency $\tilde{\omega}_m^{st} = 0$ is not a pole of the excitation coefficient. Therefore, \emph{the excitation of a static mode results in a non-resonant contribution to the scattered field}. By non-resonant, we mean that the contribution of static modes is spectrally smooth whatever the spectral range. In contrast, the contribution of other QNMs always resonates close to the corresponding eigenfrequency. Of course, in a given spectral range, QNMs with an eigenfrequency located far away in the complex plane will contribute non-resonantly. But it is important to emphasize that non-static QNMs, which are poles of the scattered field, play a different role from static QNMs, which are not poles of the scattered field.

It is meaningful from a physical point of view to separate the modal expansion of the scattered field in two distinct parts,

\begin{equation}\label{eq:Es2}
    {\bf E}_s({\bf r},\omega) = {\bf E}_s^\textrm{nr}({\bf r},\omega) + {\bf E}_s^\textrm{r}({\bf r},\omega) ,
\end{equation}

\noindent with ${\bf E}_s^\textrm{nr}$ a non-resonant term resulting from the excitation of static modes and ${\bf E}_s^\textrm{r}$ a sum of resonant terms resulting from the excitation of all other QNMs,

\begin{subequations}\label{eq:Esr_et_nr}
\begin{align}
    {\bf E}_s^\textrm{nr}({\bf r},\omega) & = \sum_p a_p^{st}(\omega) \tilde{{\bf E}}_p^{st}({\bf r}) \,, \\
    {\bf E}_s^\textrm{r}({\bf r},\omega) & = \sum_m \frac{\omega}{\omega - \tilde{\omega}_m} a_m(\omega) \tilde{{\bf E}}_m({\bf r}) \,.
\end{align}
\end{subequations}

\noindent The superscript $^{st}$ is used to distinguish the family of static modes from other QNMs. The coefficients $a_m$ and $a_m^{st}$ are related to the overlap integral between the permittivity difference $\Delta\varepsilon$, the background field and the mode. This overlap is taken at the working frequency $\omega$.

Finally, the total field can be rewritten as

\begin{equation}\label{eq:Et2}
    {\bf E}({\bf r},\omega) = \underset{\textrm{non-resonant}}{\underbrace{ {\bf E}_b({\bf r},\omega) + \sum_p a_p^{st}(\omega) \tilde{{\bf E}}_p^{st}({\bf r}) }} + \underset{\textrm{resonant}}{\underbrace{ \sum_m \frac{\omega}{\omega - \tilde{\omega}_m} a_m(\omega) \tilde{{\bf E}}_m({\bf r}) }} .
\end{equation}

\medskip

In contrast to what is suggested by Eqs.~\eqref{eq:Et} and~\eqref{eq:Es}, we can draw the following conclusions: (i) a QNM expansion of the scattered field includes a non-resonant term and (ii) the non-resonant contribution to the total field is not given by the background field alone, but results from the background field corrected by the excitation of static modes. Without the static part, the non-resonant term is erroneous and the modal expansion is not able to accurately predict all the important features of the scattering problem. In particular, the off-resonance behavior or the spectral shape of some Fano resonances cannot be correctly calculated.

In Sections 3 and 4, we illustrate with concrete examples that this conclusion is not only a formal argument. We show that the error made with a QNM expansion without static modes is significant and that static modes are of prime importance in an expansion truncated to only a few modes.

\begin{figure}[b!]
    \centering\includegraphics[width=\columnwidth]{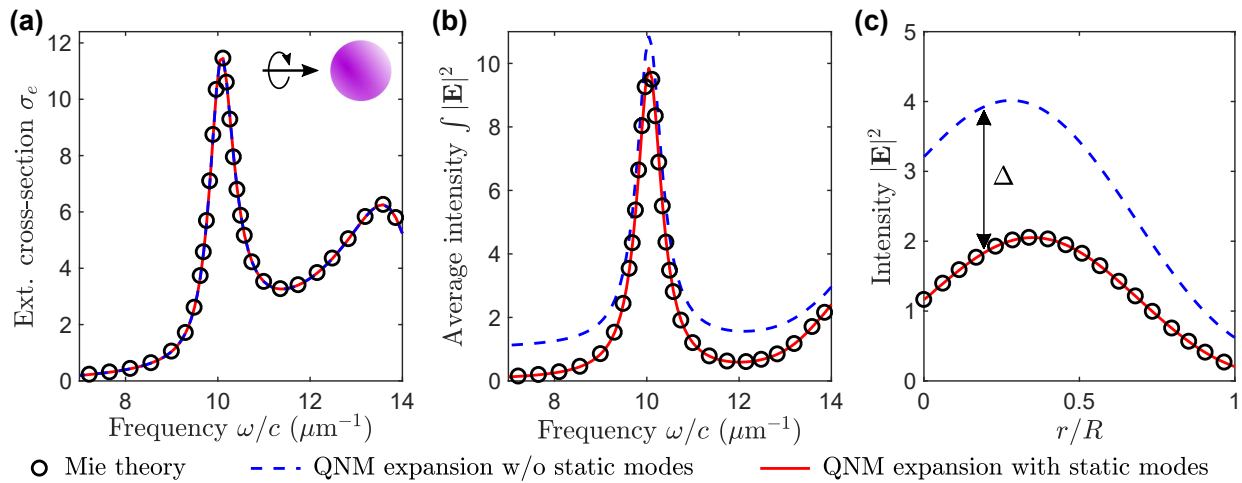}
    \caption{QNM theory applied to a lossless sphere (refractive index $n = 4$ and radius $R =75$~nm) embedded in a homogeneous medium ($n_b = 1$) and illuminated by a right-handed circularly polarized plane wave. (a) Spectrum of the extinction cross-section $\sigma_e$ normalized by the sphere apparent surface $\pi R^2$. (b) Spectrum of the average intensity enhancement inside the sphere, $(1/V_s)\int_{V_s} |\mathbf{E}|^2/|\mathbf{E}_i|^2 d^3{\bf r}$. (c) Intensity enhancement $|\mathbf{E}|^2/|\mathbf{E}_i|^2$ at $\omega/c = k_0 = 11.36$~$\mu$m$^{-1}$ along the sphere radius for $\theta = 36.23^\circ$ and $\phi = 0$. The results of a QNM expansion that includes static modes (red solid line) are compared with those of a QNM expansion without static modes (blue dashed line). Mie theory (black circles) is used as a reference. In (c), $\Delta$ represents the error made with a modal expansion without static modes. The relative error can be as large as 100\%. Modal calculations are made with 500 QNMs + 5 static modes.}
    \label{fig:lossless1}
\end{figure}

\section{Importance of static modes in the case of a lossless sphere}
\label{sec:lossless}

Let us first consider the example of light scattering by a lossless nanosphere with a constant refractive index. This simple case is interesting to underline the special role played by static modes since, in a lossless system with $\textrm{Im}(\varepsilon) = 0$, static modes do not contribute to the cross-section while they do contribute to the electromagnetic field.

A sphere (radius $R = 75$~nm and refractive index $n=4$) is embedded in a homogeneous medium ($n_b = 1$) and illuminated by a plane wave with a right circular polarization, as sketched in the inset of Fig.~\ref{fig:lossless1}(a). The incident electric field, propagating along the $-z$ direction, is given by ${\bf E}_i({\bf r}) = ({\bf e}_x + i{\bf e}_y)e^{-ik_0n_bz}$, where $({\bf e}_x,{\bf e}_y)$ are the unitary vectors of the $x$ and $y$ axis and $k_0 = \omega /c = 2\pi/\lambda$. The refractive index $n=4$ has been chosen as a rough approximation of the refractive index of silicon in the visible. Note that the choice of circular rather than linear polarization for the incident field has no impact on the conclusions and was guided only by the spherical symmetry of the system.

\subsection{Cross-section and internal field}
\label{sec:sigma_et_E}

The total field inside the sphere and the extinction cross-section $\sigma_e$ are first calculated with Mie theory, which is used as a reference, see the black circles in Fig~\ref{fig:lossless1}. Note that, since $\textrm{Im}(\varepsilon) = 0$, there is no absorption and the extinction cross-section is equal to the scattering cross-section. Figure~\ref{fig:lossless1}(a) shows the spectrum of $\sigma_e$ normalized by the sphere apparent surface $\pi R^2$ and Fig.~\ref{fig:lossless1}(b) shows the spectrum of the intensity enhancement averaged over the sphere volume, $\frac{1}{V_s}\int_{V_s} \frac{|\mathbf{E}|^2}{|\mathbf{E}_i|^2} d^3{\bf r}$. In the chosen spectral range that matches the visible range, the cross-section exhibits two resonances. In a multipole decomposition, the sharp peak close to $k_0 = 10$~$\mu$m$^{-1}$ corresponds to a magnetic dipole and the broad peak at $k_0 = 13.6$~$\mu$m$^{-1}$ corresponds to an electric dipole. Figure~\ref{fig:lossless1}(c) displays the intensity enhancement $|\mathbf{E}|^2/|\mathbf{E}_i|^2$ inside the sphere along the sphere radius for $\theta = 36.23^\circ$ and $\phi = 0$ calculated at $k_0 = 11.36$~$\mu$m$^{-1}$ between both resonances.

The results of Mie theory are compared to those of a QNM expansion with 500 modes. The total field inside the sphere is calculated with Eq.~\eqref{eq:Et2}. The extinction cross-section is calculated with the following formula,

\begin{equation}\label{eq:sigma_e}
    \sigma_e = \frac{\omega}{2P_i} \iiint_{V_s} \textrm{Im} \left [ \Delta\varepsilon {\bf E} \cdot {\bf E}_b^* \right ] d^3{\bf r} \,,
\end{equation}

\noindent where $P_i = \frac{1}{2}\varepsilon_0 c n_b |{\bf E}_i|^2$ is the Poynting vector norm of the incident plane wave and ${\bf E}$ is the total electric field calculated with the modal expansion of Eq.~\eqref{eq:Et2}.

We compare two different QNM expansions: a first one that includes 5 static modes (red solid line) in addition to 500 QNMs and a second one without static modes (blue dashed line). This second expansion amounts to set $a_p^{st} = 0$ in Eqs.~\eqref{eq:Esr_et_nr} and~\eqref{eq:Et2}. Thus, the only difference between the red solid curves and the blue dashed curves is the presence of the second non-resonant term in Eq.~\eqref{eq:Et2}. Details on the modes calculation are given in Section~\ref{sec:cal_modes}.

In Fig.~\ref{fig:lossless1}(a), the results obtained with both modal expansions are superimposed and coincide with Mie theory. It shows that, in a lossless system with $\textrm{Im}(\varepsilon) = 0$, static modes do not contribute to the cross-section. Note that it was already known that static modes do not contribute to spontaneous emission~\cite{MuljarovPRB16}. Since $\textrm{Im}(\Delta\varepsilon) = 0$ for a lossless resonator, the contribution of static modes to the cross-section reduces to the integral of $\textrm{Im} \left [ a_p^{st}(\omega) \tilde{{\bf E}}_p^{st} \cdot {\bf E}_b^* \right ]$. The quantity inside the brackets is purely real and the imaginary part is null. This is the reason why static modes do not contribute to the cross-section of a lossless sphere. We will see in Section 4 that this conclusion is not true for an absorbing system.

However, we point out that, even if static modes are not necessary to calculate the cross-section, \emph{they must be taken into account to correctly calculate the electric field inside the sphere}. Indeed, Figs.~\ref{fig:lossless1}(b) and (c) demonstrate that a QNM expansion without static modes (blue dashed curves) provides incorrect results that are not in agreement with Mie theory. Figure~\ref{fig:lossless1}(b) shows that the non-resonant background of the spectrum is not correctly calculated without static modes. The arrow labeled $\Delta$ in Fig.~\ref{fig:lossless1}(c) shows that the relative error on the intensity made with a QNM expansion without static modes can be as large as 100\%. Static modes have thus a significant impact on the field reconstruction.

\subsection{Quasinormal modes of a sphere}
\label{sec:cal_modes}

We describe here the QNMs used in the modal expansion and their calculation. The latter is based on Mie theory and amounts to calculating the zeros of an analytical function of a complex variable. More details can be found in~\cite{MuljarovPRA14,MansuPRA17,BonodPRB18,BonodJOSAB19}.

\begin{figure}[b!]
    \centering\includegraphics[width=\columnwidth]{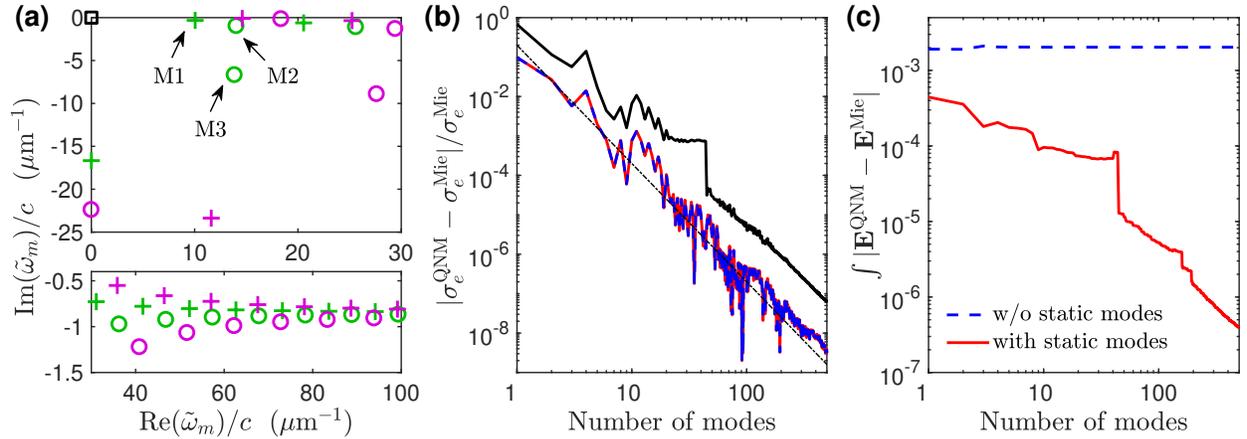}
    \caption{Convergence of the QNM expansion for the lossless sphere of Fig.~\ref{fig:lossless1}. (a) Positions of a few modes in the complex frequency plane. Modes labeled M1, M2 and M3 are responsible for the two peaks in Fig.~\ref{fig:lossless1}(a). Circles: electric (TM) modes. Crosses: magnetic (TE) modes. Black square: static modes. Modes with $l=1$ (resp. $l=2$) are shown with green (resp. purple) markers. Modes with $l>2$ are calculated but not shown here, see Supplement 1. (b) Relative error on the cross-section as a function of the number of modes. (c) Error on the total electric field integrated over the sphere volume. Results of Mie theory ($\sigma_e^\textrm{Mie}$ and ${\bf E}^\textrm{Mie}$) are taken as a reference. The convergence of a QNM expansion that includes 5 static modes (red solid line) is compared with the convergence of a QNM expansion without static modes (blue dashed line). The error is calculated at $k_0 = 9.971$~$\mu$m$^{-1}$. In (b), the black solid curve shows the mean of the relative error over the spectral range considered in Fig.~\ref{fig:lossless1} and the thin dash-dotted line displays a $1/N^3$ decrease. In (c), the QNM expansion without static modes does not converge towards the exact result provided by Mie theory. The detailed list of the 500 QNMs used in (b) and (c) can be found in Supplement 1.}
    \label{fig:lossless2}
\end{figure}

Figure~\ref{fig:lossless2}(a) displays the position in the complex frequency plane of a few eigenmodes of the lossless sphere. The three modes labeled M1, M2, and M3 are of particular interest in the spectral range considered in Fig.~\ref{fig:lossless1}. The M1 mode is responsible for the magnetic resonance close to $k_0 = 10$~$\mu$m$^{-1}$. The broad electric resonance at $k_0 = 13.6$~$\mu$m$^{-1}$ results from the superposition of the excitation of modes M2 and M3~\cite{BonodJOSAB19}.

A spherical resonator supports three distinct families of QNMs, electric modes with $H_r=0$ (also called TM modes), magnetic modes with $E_r=0$ (TE modes), and static modes (longitudinal modes) with $\tilde{\omega}_m = 0$~\cite{MuljarovPRA14}. Static modes can also be split into electric and magnetic groups. As explained in Section~\ref{sec:static_modes}, we consider only electric static modes (longitudinal electric) since magnetic static modes cannot be excited in a non-magnetic system as all their electric-field components are zero. In Fig.~\ref{fig:lossless2}(a), electric modes are shown with circles, magnetic modes are shown with crosses, and static modes are shown with squares. Furthermore, eigenmodes of a sphere are characterized by the degree $l$ (or longitudinal number) of the associated vectorial spherical harmonic, $l=1$ being a dipolar mode~\cite{BonodJOSAB19}. Eigenfrequencies that correspond to $l=1$ (resp. $l=2$) are shown with green (resp. purple) markers in Fig.~\ref{fig:lossless2}(a). Finally, each eigenfrequency is associated to $2l+1$ degenerate modes characterized by different azimuthal numbers $m \in [-l,l]$. Since we consider a right-handed circularly polarized incident field, only modes with $m=1$ can be excited. Note that QNMs with $l>2$ [not shown in Fig.~\ref{fig:lossless2}(a)] are used in the QNM expansion; the detailed list can be found in Supplement 1.

Non-static QNM frequencies are found by calculating, in the complex plane, the zeros of an analytical function whose closed-form expression involves spherical Bessel functions~\cite{MuljarovPRA14,MansuPRA17,BonodPRB18}. A different characteristic equation has to be solved for each polarization (electric and magnetic) and for each value of the degree $l \geqslant 1$. Each equation provides an infinite number of frequencies in the lower right quadrant of the complex plane, $\textrm{Re}(\tilde{\omega}_m)>0$ and $\textrm{Im}(\tilde{\omega}_m)<0$. The QNM field can then be obtained analytically from a vectorial spherical harmonic taken at the QNM complex frequency and characterized by the corresponding triplet (polarization,$l$,$m$)~\cite{MuljarovPRA14,MansuPRA17}. Modes are normalized with Eq.~\eqref{eq:Norm}. Except for a few eigenfrequencies whose real part $\textrm{Re}(\tilde{\omega}_m)$ is inside or near the spectral range of interest, most of the modes form a large-frequency tail with an almost constant imaginary part $\textrm{Im}(\tilde{\omega}_m)$, see lower inset in Fig.~\ref{fig:lossless2}(a). In addition, for each even (resp. odd) value of the degree $l$, one electric (resp. magnetic) mode has a purely imaginary eigenfrequency with $\textrm{Re}(\tilde{\omega}_m)=0$ and $\textrm{Im}(\tilde{\omega}_m) \neq 0$.

On the other hand, static modes are all degenerate with a zero frequency $\tilde{\omega}_m = 0$ whatever the values of the degree $l$ and the azimuthal number $m$, see the black square in Fig.~\ref{fig:lossless2}(a). Their field is also calculated analytically, either with specific closed-form expressions or as the limit $\tilde{\omega}_m \rightarrow 0$ of electric (TM) modes, see~\cite{MuljarovPRA14}. Since their field vanishes outside the sphere as $r \rightarrow \infty$, their normalization is not an issue; they are normalized with Eq.~\eqref{eq:Norm_sta} with an integration domain $\Omega$ in the real coordinate space.

The QNM family $(\tilde{\omega}_m,\tilde{{\bf E}}_m)$ with $\tilde{\omega}_m$ in the lower right quadrant of the complex frequency plane has to be completed by a second ``twin'' family $(-\tilde{\omega}_m^*,\tilde{{\bf E}}_m^*)$ that can easily be deduced from the first one by complex conjugation~\cite{MansuPRA17}. This property is related to the fact that causality imposes the Hermitian symmetry of the permittivity, $\varepsilon(\omega)^* = \varepsilon(-\omega^*)$. Each eigenfrequency $\tilde{\omega}_m$ in the lower right quadrant of the complex plane is paired with its $-\tilde{\omega}_m^*$ twin in the lower left quadrant, except those with a null real part. The purely imaginary frequencies and the static modes do not appear in pairs.

\subsection{Convergence of the modal expansion}

We study in this Section the convergence of the modal expansion given by Eq.~\eqref{eq:Et2} as the number of QNMs retained in the expansion is increased. Static modes are taken into account as a single non-resonant term made of the sum of the contributions of 5 static modes for $l \in [1,5]$. The convergence with this term is compared to the convergence without.

Figure~\ref{fig:lossless2}(b) shows the relative error on the cross-section $|\sigma_e^\textrm{QNM} - \sigma_e^\textrm{Mie}|/\sigma_e^\textrm{Mie}$ as a function of the number of modes. The reference value $\sigma_e^\textrm{Mie}$ is calculated with Mie theory and $\sigma_e^\textrm{QNM}$ is the cross-section calculated with Eqs.~\eqref{eq:Et2} and~\eqref{eq:sigma_e}. The error made with a QNM expansion that includes static modes (red solid curve) is compared to the error made with an expansion that omits static modes (blue dashed curve). Since static modes do not contribute to the cross-section in this lossless system, both curves are superimposed.

The red and blue curves correspond to the relative error calculated at the frequency of the magnetic resonance, $k_0 = 9.971$~$\mu$m$^{-1}$. Modes are sorted in descending order according to their contribution to the cross-section, $|\iiint_{V_s} \textrm{Im} \left [ \Delta\varepsilon \alpha_m \tilde{{\bf E}}_m \cdot {\bf E}_b^* \right ] d^3{\bf r}|$. The first three modes are, by descending order of importance, M1, M2, and M3. The relative error decreases as $1/N^3$, as shown by the thin dash-dotted line, to reach a value smaller than $10^{-8}$ when 500 modes are retained in the QNM expansion. The 500 QNMs considered in Fig.~\ref{fig:lossless2}(b) are composed of 288 modes with $l=1$, 138 modes with $l=2$, 54 modes with $l=3$, 18 modes with $l=4$, and 2 modes with $l=5$. These numbers include modes with $\textrm{Re}(\tilde{\omega}_m)>0$ and their twins at $-\tilde{\omega}_m^*$. The detailed list of the 500 QNMs and their sorting are given in Supplement 1. We have checked that it is possible to further decrease the error by increasing the number of modes.

The black solid curve in Fig.~\ref{fig:lossless2}(b) displays the mean of the relative error over the spectral range considered in Fig.~\ref{fig:lossless1} calculated as $\frac{1}{M}\sum_{i=1}^M \textrm{Err}_i$, with $\textrm{Err}_i = |\sigma_e^\textrm{QNM}(\omega_i) - \sigma_e^\textrm{Mie}(\omega_i)|/\sigma_e^\textrm{Mie}(\omega_i)$ and $M = 100$ the number of frequency points $\omega_i$ used in Fig.~\ref{fig:lossless1}. Note that the QNMs are sorted at $k_0 = 9.971$~$\mu$m$^{-1}$ and the ranking is kept the same for each frequency.

We have seen in Section~\ref{sec:sigma_et_E} that static modes do not contribute to the cross-section but are essential for the field reconstruction. Therefore, in addition to the relative error on the cross-section, we have calculated at $k_0 = 9.971$~$\mu$m$^{-1}$ the error on the total field inside the sphere, $|{\bf E}^\textrm{QNM}({\bf r}) - {\bf E}^\textrm{Mie}({\bf r})|$. This error depends on the point under consideration. To obtain a global, position independent, quantity, we have integrated the error over the sphere volume. The result is shown in Fig.~\ref{fig:lossless2}(c). The error made with a modal expansion that includes static modes (red solid curve) decreases as the number of modes is increased, showing the convergence of the QNM expansion. On the contrary, \emph{the modal expansion without static modes (blue dashed curve) does not converge towards the exact result provided by Mie theory}. It is noteworthy that, even with a single QNM, the result provided by taking into account the static non-resonant contribution is significantly better. Adding more and more modes does not improve the incorrect result provided by a modal expansion without static modes; the error remains constant. The 500 QNMs used in the modal expansion and their ranking are the same as in Fig.~\ref{fig:lossless2}(b).

\section{Importance of static modes in the case of an absorbing and dispersive sphere}

The results in Section~\ref{sec:lossless} reveal that static modes do not contribute to the extinction cross-section of a lossless sphere but are crucial to correctly reconstruct the internal field. We show in this Section that, in the case of an absorbing and dispersive system, the important role played by static modes in the field reconstruction reflects into the calculation of the extinction and absorption cross-sections.

We consider the same geometry as in Figs.~\ref{fig:lossless1} and~\ref{fig:lossless2}, a sphere of radius $R =75$~nm embedded in a medium with $n_b = 1$, but we replace the constant refractive index inside the sphere by an absorbing and dispersive medium. Its permittivity is given by a Lorentz model, $\varepsilon(\omega) = \varepsilon_\infty \left (1 - \frac{\omega_p^2}{\omega^2 - \omega_0^2 + i\omega\gamma} \right )$, with $\varepsilon_\infty = 8.51$, $\omega_p = 3.62 \times 10^{15}$~rad.s$^{-1}$, $\omega_0 = 5.09 \times 10^{15}$~rad.s$^{-1}$, and $\gamma = 1.16 \times 10^{14}$~rad.s$^{-1}$. The parameters of the Lorentz model are chosen to fit the data of silicon tabulated in~\cite{Green} in the $[400-800]$~nm spectral range.

The incident beam is the same circularly polarized plane wave as in Figs.~\ref{fig:lossless1} and~\ref{fig:lossless2}. The incident electric field, propagating along the $-z$ direction, is given by ${\bf E}_i({\bf r}) = ({\bf e}_x + i{\bf e}_y)e^{-ik_0n_bz}$, where $({\bf e}_x,{\bf e}_y)$ are the unitary vectors of the $x$ and $y$ axis and $k_0 = \omega /c = 2\pi/\lambda$.

\begin{figure}[b!]
    \centering\includegraphics[width=\columnwidth]{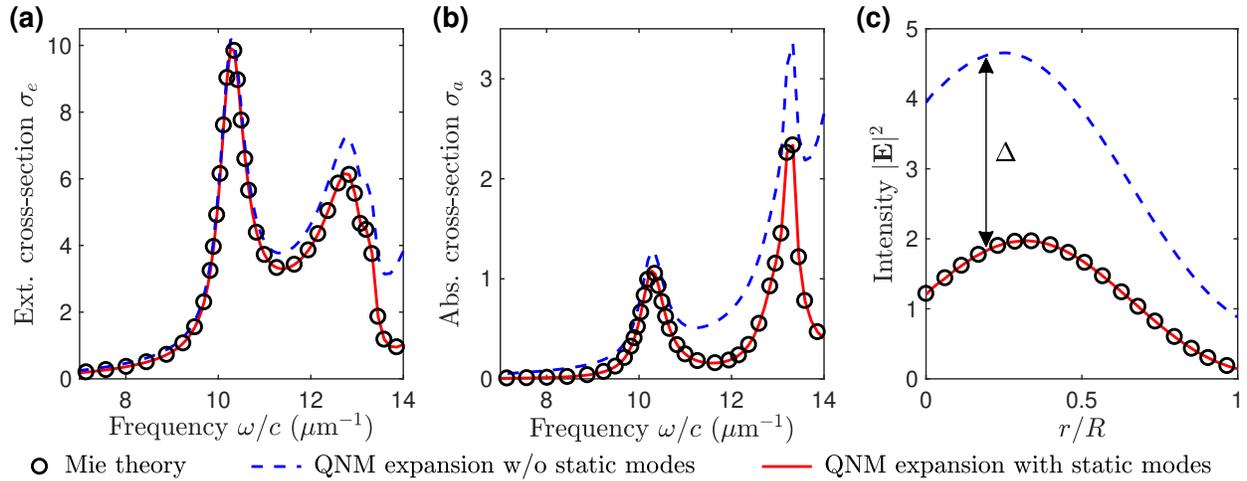}
    \caption{QNM theory applied to an absorbing and dispersive sphere (silicon, radius $R =75$~nm) embedded in a homogeneous medium ($n_b = 1$) and illuminated by a right-handed circularly polarized plane wave. (a) Spectrum of the extinction cross-section $\sigma_e$. (b) Spectrum of the absorption cross-section $\sigma_a$. (c) Intensity enhancement $|\mathbf{E}|^2/|\mathbf{E}_i|^2$ at $\omega/c = k_0 = 11.36$~$\mu$m$^{-1}$ along the sphere radius for $\theta = 36.23^\circ$ and $\phi = 0$. The results of a QNM expansion that includes static modes (red solid line) are compared with those of a QNM expansion without static modes (blue dashed line). Mie theory (black circles) is used as a reference. In (b) and (c), the cross-sections are normalized by the sphere apparent surface $\pi R^2$. In (c), $\Delta$ represents the error made with a modal expansion without static modes. The relative error can be larger than 100\%. Modal calculations are made with 650 QNMs + 5 static modes.}
    \label{fig:lossy1}
\end{figure}

\subsection{Cross-section and internal field}

The total field inside the sphere, the extinction cross-section $\sigma_e$, and the absorption cross-section $\sigma_a$ are first calculated with Mie theory, which is used as a reference, see the black circles in Fig~\ref{fig:lossy1}. Figures~\ref{fig:lossy1}(a) and (b) show the spectra of $\sigma_e$ and $\sigma_a$ normalized by the sphere apparent surface $\pi R^2$. Similarly to the lossless sphere, the absorbing sphere exhibits one magnetic resonance ($k_0 = 10.3$~$\mu$m$^{-1}$), which is due to the excitation of a single QNM, and one electric resonance ($k_0 \approx 12.8$~$\mu$m$^{-1}$), which results from the excitation of several QNMs. Figure~\ref{fig:lossy1}(c) displays the intensity enhancement $|\mathbf{E}|^2/|\mathbf{E}_i|^2$ inside the sphere along the sphere radius for $\theta = 36.23^\circ$ and $\phi = 0$ calculated at $k_0 = 11.36$~$\mu$m$^{-1}$ between both resonances.

The results of Mie theory are compared to those of a QNM expansion with 650 modes. The total electric field inside the sphere is calculated with Eq.~\eqref{eq:Et2}. The extinction cross-section is calculated with Eq.\eqref{eq:sigma_e} and the absorption cross-section is calculated as

\begin{equation}\label{eq:sigma_a}
    \sigma_a = \frac{\omega}{2P_i} \iiint_{V_s} \textrm{Im}(\varepsilon) |{\bf E}|^2 d^3{\bf r} \,,
\end{equation}

\noindent where $P_i = \frac{1}{2}\varepsilon_0 c n_b |{\bf E}_i|^2$ is the Poynting vector norm of the incident plane wave and ${\bf E}$ is the total electric field calculated with the modal expansion of Eq.~\eqref{eq:Et2}.

We compare two different QNM expansions: a first one that includes 5 static modes (for $l \in [1,5]$, red solid line) in addition to 650 QNMs and a second one without static modes (blue dashed line). This second expansion amounts to set $a_p^{st} = 0$ in Eqs.~\eqref{eq:Esr_et_nr} and~\eqref{eq:Et2}. Thus, the only difference between the red solid curves and the blue dashed curves is the presence of the second non-resonant term in Eq.~\eqref{eq:Et2}.

In Figs.~\ref{fig:lossy1}(a) and (b), the results obtained with both modal expansions are different, especially at high frequencies when absorption in silicon is larger. Only the modal expansion that includes static modes (red solid curves) is in agreement with Mie theory. The blue dashed curves indicate that \emph{omitting static modes provides incorrect results}. At high frequencies, the relative error can even exceed 100\%. In contrast to the case of a lossless system, static modes do contribute to the extinction cross-section, as well as to the absorption cross-section. The latter is a measure of the average intensity inside the sphere.

Finally, Fig.~\ref{fig:lossy1}(c) illustrates the significant error on the total field inside the sphere that is made with a QNM expansion without static modes (blue dashed curve). The arrow labeled $\Delta$ in Fig.~\ref{fig:lossy1}(c) shows that the relative error on the intensity can be larger than 100\%. Static modes are therefore crucial to reconstruct the field correctly.

\begin{figure}[t!]
    \centering\includegraphics[width=\columnwidth]{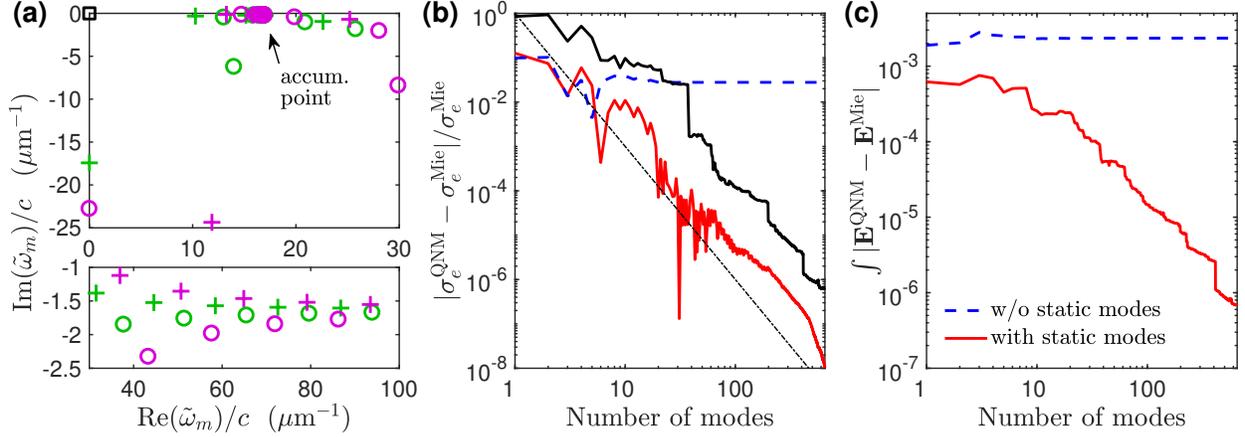}
    \caption{Convergence of the QNM expansion for the absorbing sphere of Fig.~\ref{fig:lossy1}. (a) Positions of a few modes in the complex frequency plane. Circles: electric (TM) modes. Crosses: magnetic (TE) modes. Black square: static modes. Modes with $l=1$ (resp. $l=2$) are shown with green (resp. purple) markers. Modes with $l>2$ are calculated but not shown here, see \textbf{SI}. The accumulation point corresponds to the pole of the permittivity. (b) Relative error on the extinction cross-section as a function of the number of modes. (c) Error on the total electric field integrated over the sphere volume. Results of Mie theory ($\sigma_e^\textrm{Mie}$ and ${\bf E}^\textrm{Mie}$) are taken as a reference. The convergence of a QNM expansion that includes 5 static modes (red solid line) is compared with the convergence of a QNM expansion without static modes (blue dashed line). The error is calculated at $k_0 = 10.267$~$\mu$m$^{-1}$. In (b), the black solid curve shows the mean of the relative error over the spectral range considered in Fig.~\ref{fig:lossy1} and the thin dash-dotted line displays a $1/N^3$ decrease. In (b) and (c), the QNM expansion without static modes does not converge towards the exact result provided by Mie theory. The detailed list of the 650 QNMs used in (b) and (c) can be found in Supplement 1.}
    \label{fig:lossy2}
\end{figure}

\subsection{Convergence of the modal expansion}

We study in this Section the convergence of the modal expansion given by Eq.~\eqref{eq:Et2} as the number of QNMs retained in the expansion is increased. Static modes are taken into account as a single non-resonant term made of the sum of the contributions of 5 static modes for $l \in [1,5]$. The convergence with this term is compared to the convergence without.

Figure~\ref{fig:lossy2}(a) displays the position in the complex frequency plane of a few eigenmodes. Eigenfrequencies that correspond to $l=1$ (resp. $l=2$) are shown with green (resp. purple) markers. Electric (TM) modes are shown with circles, magnetic (TE) modes are shown with crosses, and static modes are shown with squares. Modes with $l>2$ are calculated but not shown here; the detailed list of the 650 QNMs used in the modal expansion can be found in Supplement 1. The distribution of the eigenfrequencies in the complex plane is very similar to the case of the lossless sphere. The main difference is the presence of an accumulation point: for each polarization and each value of $l$, the eigenfrequencies accumulate towards the pole of the permittivity~\cite{MansuPRA17}.

Figure~\ref{fig:lossy2}(b) shows the relative error on the extinction cross-section $|\sigma_e^\textrm{QNM} - \sigma_e^\textrm{Mie}|/\sigma_e^\textrm{Mie}$ as a function of the number of modes. The reference value $\sigma_e^\textrm{Mie}$ is calculated with Mie theory and $\sigma_e^\textrm{QNM}$ is the cross-section calculated with Eqs.~\eqref{eq:Et2} and~\eqref{eq:sigma_e}. The error made with a QNM expansion that includes static modes (red solid curve) is compared to the error made with an expansion that omits static modes (blue dashed curve). The error is calculated at $k_0 = 10.267$~$\mu$m$^{-1}$ and the modes are sorted in descending order according to their contribution to the extinction cross-section. The modal expansion without static modes is not correct since it does not converge towards the results of Mie theory. On the contrary, the modal expansion that includes 5 static modes converges as $1/N^3$, as shown by the thin dash-dotted line, and reaches an error as small as $10^{-8}$ when 650 modes are retained in the QNM expansion. The 650 QNMs are composed of 319 modes with $l=1$, 222 modes with $l=2$, 88 modes with $l=3$, 19 modes with $l=4$, and 2 modes with $l=5$. These numbers include modes with $\textrm{Re}(\tilde{\omega}_m)>0$ and their twins at $-\tilde{\omega}_m^*$. The detailed list of the 650 QNMs and their sorting are given in Supplement 1.

The black solid curve in Fig.~\ref{fig:lossy2}(b) displays the mean of the relative error over the spectral range considered in Fig.~\ref{fig:lossy1} calculated as $\frac{1}{M}\sum_{i=1}^M \textrm{Err}_i$, with $\textrm{Err}_i = |\sigma_e^\textrm{QNM}(\omega_i) - \sigma_e^\textrm{Mie}(\omega_i)|/\sigma_e^\textrm{Mie}(\omega_i)$ and $M = 100$ the number of frequency points $\omega_i$ used in Fig.~\ref{fig:lossy1}. The QNM expansion includes the same 5 static modes as the red solid curve. Note that the QNMs are sorted at $k_0 = 10.267$~$\mu$m$^{-1}$ and the ranking is kept the same for each frequency.

Figure~\ref{fig:lossy2}(c) displays the error on the total field inside the sphere calculated at $k_0 = 10.267$~$\mu$m$^{-1}$. The point-dependent error $|{\bf E}^\textrm{QNM}({\bf r}) - {\bf E}^\textrm{Mie}({\bf r})|$ is integrated over the sphere volume to obtain a global, position-independent, quantity. The error made with a modal expansion that includes static modes (red solid curve) decreases as the number of modes is increased, showing the convergence of the QNM expansion. On the contrary, the modal expansion without static modes (blue dashed curve) does not converge towards the exact result provided by Mie theory. It is noteworthy that, even with a single QNM, the result provided by taking into account the static non-resonant contribution is significantly better. Adding more and more modes does not improve the incorrect result provided by a modal expansion without static modes. The 650 QNMs used in the modal expansion and their ranking are the same as in Fig.~\ref{fig:lossy2}(b).

\section{Conclusion and perspectives}

We have shown that, in the case of resonators made of absorbing dielectric materials, static modes (eigenmodes with zero eigenfrequency) play a key role in QNM theory. The excitation of static modes provides a non-resonant contribution to the modal expansion of the scattered field. In this work, the adjective ``non-resonant'' is used to characterize something that is spectrally smooth whatever the spectral range, i.e., that does not resonate, even outside the spectral range of interest. The non-resonant term resulting from the excitation of static modes is added to the background field to build a non-resonant contribution to the total field. Therefore, static modes play a crucial physical role since they largely contribute to the off-resonance signal to which resonances are added in amplitude, possibly leading to interference phenomena and Fano resonances.

The importance of static modes in the field reconstruction has been quantified in the case of a sphere. A QNM expansion that omits static modes does not converge at all towards the exact result provided by Mie theory; it results in significant relative errors that can even exceed 100\%. In the case of a lossless dielectric material, static modes contribute to the internal field but not to the cross-section. In the case of an absorbing material, static modes contribute to the internal field and to all cross-sections, absorption, scattering, and extinction.

This work on a simple, analytically solvable, case is a proof of principle that demonstrates the importance of static modes. It paves the way for a more systematic study of the role of static modes in more complex geometries for which QNMs are calculated numerically~\cite{WeiPRB2018,QNMSolver}. From a physical perspective, it would be interesting to relate the non-resonant contribution to the modal expansion derived from static modes to the one derived from causality arguments in~\cite{BonodPRB18,BonodJOSAB19}.

\section*{Funding}
This work was supported by the French National Agency for Research (ANR) under the project "Resonance" (ANR-16-CE24-0013).

\section*{Acknowledgments}
The author warmly thanks Jean-Paul Hugonin for stimulating discussions and for providing a Mie theory code.

\medskip
\noindent See Supplement 1 for supporting content.

%%%%%%%%%%%%%%%%%%%%%%% References %%%%%%%%%%%%%%%%%%%%%%%%%

%%%%%%%%%% If using BibTeX:
%\bibliography{QNMstaticRefs}

%

\end{document}